\documentclass[dvips,12pt]{article}
\usepackage{epsf,graphics}
\usepackage{a4wide}
\usepackage{amsmath}
\usepackage{latexsym}
\begin{document}

\def\d{{\rm d}}
\def\reg{{\rm reg}}

\def\veck{{\pmb{k}}}
\def\vecp{{\pmb{p}}}
\def\vecv{{\pmb{v}}}

\def\vecE{{\pmb{E}}}
\def\vecF{{\pmb{F}}}
\def\vecA{{\pmb{A}}}

\def\vecPi{{\pmb{\Pi}}}
\def\vecPir2{{\pmb{\Pi}}_r^2}

\def\vecrho{{\pmb{\rho}}}
\def\vecvarphi{{\pmb{\varphi}}}
\def\vecnabla{{\pmb{\nabla}}}

\newcommand{\Atop}[2]{\genfrac{}{}{0pt}{}{#1}{#2}}
\thispagestyle{empty}
 \begin{flushright}
 {\tt University of Bergen, Department of Physics}    \\[2mm]
 {\tt Scientific/Technical Report No.1996-03}    \\[2mm]
 {\tt ISSN 0803-2696} \\[5mm]
 {atom-ph/9604007} \\[5mm]
 {April 1996}           \\
\end{flushright}
 \vspace*{3cm}

 \begin{center}
 {\bf \large Interaction of a slow monopole
with a hydrogen atom}
 \vspace{1cm}

{\bf Ya.M.Shnir{\footnote{On leave of absence from the 
Institute of Physics, Academy of
Sciences of Belarus, Minsk, Belarus}}}\\[3pt]
{\it Department of Mathematics,}\\
{\it Technical University of Berlin, Germany}\\[2pt]
and \\[2pt]
{\it  Department of Physics, University of Bergen,}\\ 
   {\it All\'egt.~55, N-5007 Bergen, Norway}

\end{center}
\begin{abstract}
The electric dipole moment of the hydrogen-like atom induced by  
a monopole moving outside the electron shell is calculated. 
The correction to the energy of the ground state of the  
hydrogen atom due to this interaction is calculated.
\end{abstract}

\medskip

As is well known \cite{1} the interaction of a monopole with the
atoms of  matter  is  fundamental  to  the  principle  of operation of
a  number  of  detectors  (scintillators,  plastics, emulsions, etc.)
used in  the  experimental  search  for  magnetic charges.  In  this
connection the interaction of a monopole with an atom has been fairly 
well studied \cite{2}--\cite{4}. Unfortunately, for the process of
ionization of an atom by a monopole, there is a large natural background, 
which makes it difficult to identify genuine monopole tracks \cite{5}.
Therefore searches are continuing for effects that make it possible 
to distinguish a monopole from a heavy nucleus in interaction with atoms.

For example, in Ref.6 an estimate was obtained for the probability 
of excitation of an atom by a monopole on the basis of 
a quantum-mechanical analog of the Callan-Rubakov effect that occurs 
in the passage of a monopole through an electron. 
Despite its specific character, this process is obviously not dominant 
in the overall picture of the interaction of a monopole with an atom.

It was noted in Ref.4, that the interaction of a monopole with an atom 
is also characterized by a special spatial asymmetry associated with 
the space parity nonconservation in the theory with magnetic charge  
\cite{7}, \cite{8}. These effects have been studied for the model 
example of  a charge-dyon bound system \cite{9}--\cite{12}.

In the present note we calculate the electric dipole moment of 
the hydrogen-like atom induced by  a monopole moving outside the 
electron shell (see Fig.1) as well as the correction to the ground 
state energy of a hydrogen-like atom.
\begin{figure}[htb]
\begin{center}
\setlength{\unitlength}{1cm}
\begin{picture}(16,8.0)
\put(0.5,-2.0)
{\mbox{\epsfysize=9.0cm\epsffile{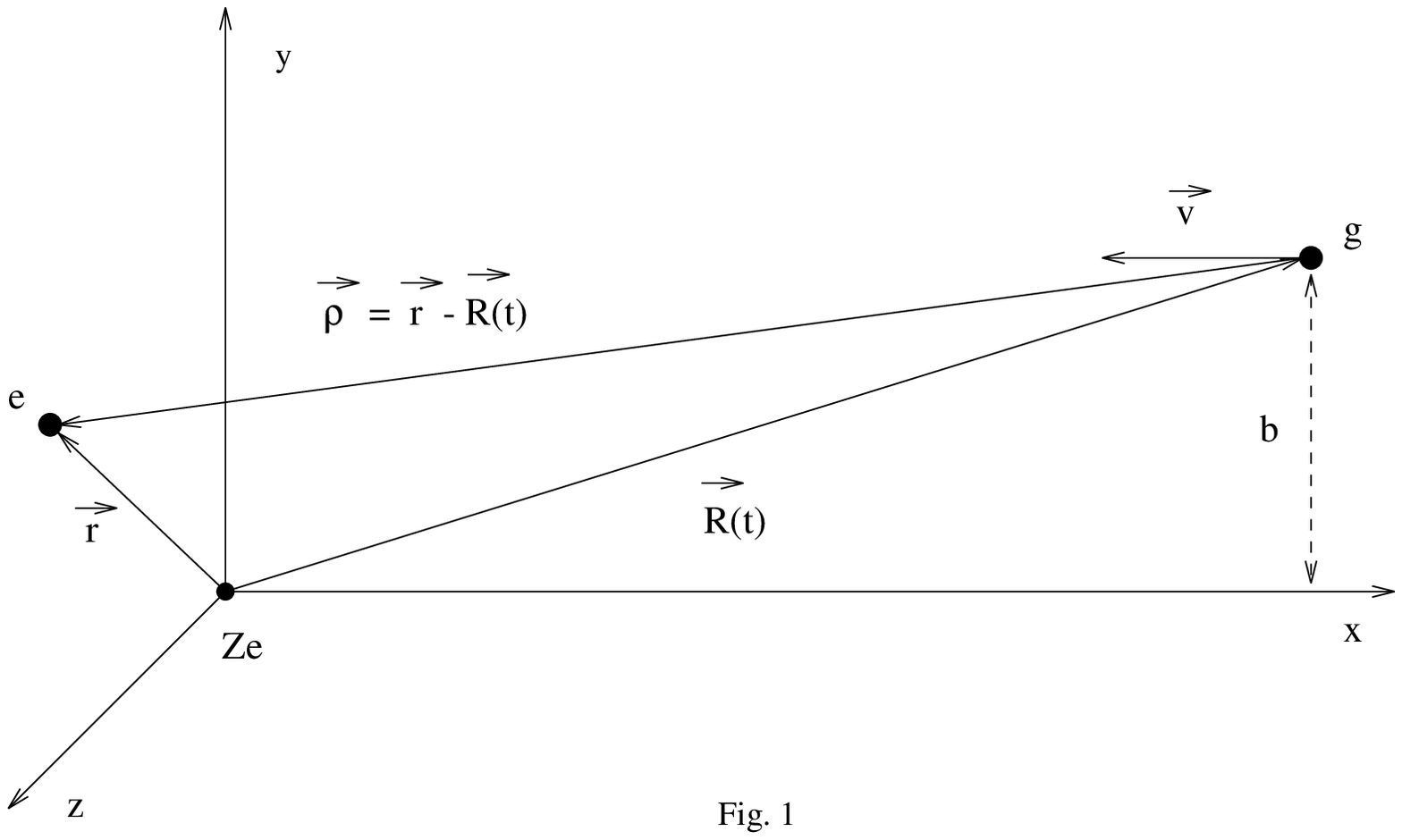}}}
\end{picture}
\vspace*{20mm}
\end{center}
\end{figure}

Suppose that a nucleus with charge $Q=Ze$  is at rest at the origin, 
and that a monopole with a magnetic charge $g$ is incident on it 
with impact parameter $b$ and constant velocity ${\bf V} = - {\bf i}v$. 
Let 
$$
{\bf R}(t) = {\bf i} vt + {\bf j}b
$$
be the coordinates of the monopole in the system associated 
with the nucleus,  $m$  the electron mass,
$$
{\bf r} = {\bf i}x + {\bf j}y + {\bf k}z 
$$
the electron coordinates, and  
$$\vecrho = {\bf r} - {\bf R} = (x-vt){\bf i} + (y-b){\bf j} +
z{\bf k}, $$a vector directed from the monopole to the electron. 
Furthermore, we define
 $$\varphi = ~{\rm actan}~\frac{\rho_y}{\rho_x}\quad {\rm and}~ ~\theta =
~{\rm arccos}~\frac{\rho_z}{\rho}
$$
 as the azimuthal and polar angles of the electron in the coordinate 
system moving together with the monopole.

 In a first approximation, we can assume that the monopole 
only excites the electron states, but does not disturb 
the relatively massive nucleus. 
In order to calculate the probability of the corresponding transitions, 
we note that the Hamiltonian operator describing this system 
(here and in what follows, $\hbar = c = 1$) is
\begin{eqnarray}                        \label{Hamiltonian}
H&=&\frac {1}{2m}(\vecnabla - ie \vecA^D)^2 - \frac{eQ}{r} 
\nonumber\\
&=&\frac {1}{2m} {\vecPi_r}^2+ \frac {{\bf L}^2}{2mr^2}  -
\frac{eQ}{r} + W_1 + W_2,
\end{eqnarray}
where 
$$\vecA^{D} =
g\frac{(1-\cos \theta )} {\rho \sin \theta } {\hat {\vecvarphi}}$$
 is the Dirac monopole potential, $\vecPi = \vecnabla - ie \vecA^D$,
${\vecPi_r} = 1/2 ({\hat {\bf r}} \vecPi + \vecPi {\hat {\bf r}} )$.
 Here we take into account that 
$ (\vecnabla \vecA^{D}) = (\vecrho \vecA^{D}) = 0$ 
and used the definition of  the operator of  total angular momentum 
of the electron,
${\bf L} = [{\bf r}, \vecPi] $.
 
The operator (1)  differs from the Hamiltonian of a hydrogen-like 
atom by the terms
\begin{eqnarray}                        \label{Perturb}
W_1 &=& \frac{e}{m} ({\vecA}^D \vecnabla) 
= \frac{i\mu}{m \rho^2(1 + \cos \theta)} \frac{\partial}{\partial \varphi},
\nonumber\\
W_2 &=& \frac{e^2}{2m} (\vecA^D)^2 
= \frac{\mu^2 }{2m \rho^2} \frac {\sin ^2 \theta}{(1 + \cos \theta)^2},
\end{eqnarray}
which describe the interaction of the atomic electron 
with the monopole (here $\mu = eg$). 

If the monopole passes sufficiently far from the atom, 
so that $\varepsilon = {r}/{R}\ll 1$, the probability of a transition 
of the electron from the initial state $|n>$ to a state $|m> $ 
is determined by the matrix element
$<m|W|n> $, where $W = W_1 + W_2$ can be regarded as a perturbation operator.

Now we take into account that for small $\varepsilon$
\begin{equation}
\cos \theta \approx \frac{z}{R} + O(\varepsilon ^2), \qquad
\frac{1}{\rho ^2} \approx \frac {1}{R^2} + \frac {1}{R^4} (2vtx + 2by) 
+O(\varepsilon ^3)
\end{equation}
and hence
\begin{eqnarray}                        \label{Perturb-exp}
W_1&\approx&\frac{\mu}{mR^2} \biggl((b-y)P_x + (x - vt)P_y\biggr) 
+O(\varepsilon ^2) = 
\nonumber\\
&=& \frac{\mu}{mR^2}\biggl(iL_z + (bP_x -vtP_y)\biggr) +O(\varepsilon ^2) ;
\nonumber\\
W_2&=&\frac{\mu^2}{2mR^4} \biggl((b-y)^2 + (x - vt)^2\biggr) 
+O(\varepsilon ^3) =  
\nonumber\\
&=&\frac{\mu^2}{2mR^2} + \frac{\mu^2}{mR^4}\biggl( \frac{(x^2 + y^2)}{2} - 
(vtx+ by) \biggr) +O(\varepsilon ^3).                  
\end{eqnarray}

It is worth noting, that the operator $W$ Eq.(2) represents 
a mixture of a scalar and a pseudoscalar. It means, 
that among dipole transitions in the hydrogen
atom spectrum, stipulated by external monopole perturbation, there are
transitions which violate  parity and strictly forbidden in the
usual case. Indeed, let us consider the correction to the wave function 
of the ground state of the hydrogen-like atom 
$|\Psi_0>  \equiv |1,0,0 > = R_{10}(r) Y_{00} =
e^{-r}/\sqrt{\pi}$ due to the perturbation:
 \begin{equation}                        \label{pert-func}
|{\tilde \Psi}_0> 
= |\Psi_0> + \sum_{n,l,m}^{}\frac{<n,l,m|W|1,0,0>}{E_n - E_0}
|n,l,m>
\end{equation}
where $n,l,m$ are the usual principal, orbital and magnetic quantum numbers,
and the energy $E_n = m{Q^2e^2}/{n^2}$. 

Using the standard definition of the spherical harmonics $Y_{lm}$ 
(see e.g. \cite{13}), one can write
\begin{equation}
x = r \sqrt{\frac{2\pi}{3}}(Y_{11} - Y_{1-1});\quad 
y= -i  r \sqrt{\frac{2\pi}{3}}(Y_{11} + Y_{1-1});
\end{equation}
$$ 
x^2 + y^2 = \frac{2}{3}r^2 \left( 1 + {\sqrt{\frac{4\pi}{5}}} Y_{20}\right).
$$
So, the action of the perturbation operator (\ref{Perturb}) 
on the ground state wave function gives
\begin{eqnarray}                        \label{ground-pert}
W_1 |\Psi_0>&\approx&\frac{\mu}{{\sqrt 6} mR^2} R_{10} \left((ivt + b)Y_{11} +
(ivt -b)Y_{1-1}\right);
\nonumber\\
W_2 |\Psi_0>&\approx&\frac{\mu^2}{2mR^2} 
R_{10}Y_{00} - \frac{\mu^2}{{\sqrt 6} mR^4}  r R_{10} 
\left((vt - ib)Y_{11} - (vt + ib) Y_{1-1} \right)+
\nonumber\\
&+&\frac{\mu^2}{3mR^4} r^2 R_{10} \left( Y_{00} + \frac{1}{{\sqrt 5}} Y_{20}
\right).
\end{eqnarray}  

Thus, we have nonzero matrix elements of the perturbation operator
\begin{eqnarray}                        \label{matrix-pert}
<n,0,0|W|\Psi_0>&\approx&\frac{\mu^2}{2mR^2} I_n[2] 
+ \frac{\mu^2}{3mR^4} I_n[4] + O(\varepsilon^3);
\nonumber\\
<n,2,0|W|\Psi_0>&\approx&\frac{\mu^2}{3{\sqrt 5}mR^4} I_n[4] 
+ O(\varepsilon^3);
\nonumber\\
<n,1,1|W|\Psi_0>&\approx&\frac{\mu}{{\sqrt 6} mR^2} 
\biggl[ I_n[2](b + ivt) +
 \frac{\mu}{R^2} I_n[3] (vt - ib) \biggr] + O(\varepsilon^3);
 \nonumber\\
<n,1,-1|W|\Psi_0>&\approx&\frac{\mu}{{\sqrt 6} mR^2} 
\biggl[ I_n[2](ivt - b) +
 \frac{\mu}{R^2} I_n[3] (ib + vt) \biggr]+ O(\varepsilon^3),
 \end{eqnarray}  
where the radial integrals are 
$$
I_n[k] \equiv  \int\limits_{0}^{\infty}dr r^k R_{n0} R_{10}$$
In particular, using these expressions, one can  calculate the first order
correction to the energy of the hydrogen atom ground state 
 \begin{equation}          \label{delta-en}
\Delta E_0 = <\Psi_0|W|\Psi_0> \approx\frac{\mu^2}{2m R^2}
+  O(\varepsilon^{3}).
\end{equation}
This result seems to be quite natural. 
Inded, one can write (\ref{delta-en}) as $\Delta E_0 \approx k H$, where $
k = e^2 g/2m \sim e/2m$ is the classical magnetic moment 
of the hydrogen atom and $H = g/R^2$ is the Coulomb magnetic field 
of a monopole (here we used the charge quantization condition). 

Taking into account the Eqs.(\ref{matrix-pert}), we can write 
the expression for the perturbed ground state wave function 
\begin{eqnarray}                        \label{ground-pert-func}
|{\tilde \Psi}_0> 
= |\Psi_0>&+&\sum_{n=1}^{\infty} \frac{<n,0,0|W|1,0,0>}{E_n - E_0}|n,0,0> 
+ \sum_{n=2}^{\infty} \frac{<n,1,1|W|1,0,0>}{E_n - E_0}|n,1,1>  
\nonumber\\
&+&\sum_{n=2}^{\infty} \frac{<n,1,-1|W|1,0,0>}{E_n - E_0}|n,1,-1> + 
 \sum_{n=2}^{\infty} \frac{<n,2,0|W|1,0,0>}{E_n - E_0}|n,2,0> 
\nonumber\\
= |\Psi_0> &+&\frac{\mu^2}{2mR^2} 
\sum_{n=1}^{\infty}\frac{I_n[2]}{E_n - E_0}|n,0,0> 
+  \frac{\mu^2}{3mR^4} \sum_{n=1}^{\infty}\frac{I_n[4]}{E_n - E_0}|n,0,0> 
\nonumber\\
&+&\frac{\mu}{{\sqrt 6} mR^2} 
\sum_{n=2}^{\infty} \frac{1}{E_n -E_0} \left(I_n[2](b - ivt) 
+ \frac{\mu}{R^2}  I_n[3] (vt - ib) \right) |n,1,1> 
\nonumber\\
 &+&\frac{\mu}{{\sqrt 6} mR^2} 
\sum_{n=2}^{\infty} \frac{1}{E_n -E_0} \left(I_n[2](ivt -b ) 
+ \frac{\mu}{R^2}  I_n[3] (vt + ib) \right) |n,1,-1> 
\nonumber\\
 &+&\frac{\mu^2}{3{\sqrt 5}mR^2} 
\sum_{n=2}^{\infty} \frac{I_n[4]}{E_n - E_0}|n,2,0> + O(\varepsilon^{3}).
\end{eqnarray}
 It is clear from Eq.(\ref{ground-pert-func}) that in the presence 
of a monopole
the hydrogen-like atom has nonzero electric dipole moment
\begin{eqnarray}                        \label{dipole-mom}
d = e<{\tilde \Psi}_0|{\bf r}|{\tilde \Psi}_0> 
= e <\Psi_0|{\bf r}|\Delta \Psi_0> + e<\Delta \Psi_0|{\bf r}|\Psi_0>
\end{eqnarray}
Indeed, taking into account (\ref{ground-pert-func}), one can obtain
\begin{eqnarray}                        \label{dipole-mom-matr}
<\Psi_0|z|\Delta\Psi_0>&=&0;
\nonumber\\
<\Psi_0|x + iy|\Delta\Psi_0>
&\approx&\frac{\mu}{3mR^2}(ivt -b)\sum_{n=2}^{\infty} 
\frac{I_n[3]I_n[2]}{E_n -E_0}  
\nonumber\\
&+&\frac{\mu^2}{3mR^4}(vt + ib) 
\sum_{n=2}^{\infty} \frac{(I_n[3])^2}{E_n -E_0} + O(\varepsilon^{3});
 \nonumber\\
<\Psi_0|x -iy|\Delta\Psi_0>&
\approx&\frac{\mu}{3mR^2}(ivt + b)\sum_{n=2}^{\infty} 
\frac{I_n[3]I_n[2]}{E_n -E_0} 
\nonumber\\
&+&\frac{\mu^2}{3mR^4}(vt - ib) 
\sum_{n=2}^{\infty} \frac{(I_n[3])^2}{E_n -E_0} + O(\varepsilon^{3}).
\end{eqnarray}
Then we have
\begin{equation}                        \label{dip-fin}
 <{\tilde \Psi}_0|z|{\tilde \Psi}_0> =0;
\end{equation}
 \begin{eqnarray}  
<{\tilde \Psi}_0|x+iy|{\tilde \Psi}_0>
&=&- <{\tilde \Psi}_0|x-iy|{\tilde \Psi}_0>  
\nonumber\\
&=&-\frac{2\mu b}{3mR^2} \sum_{n=2}^{\infty}
 \frac{I_n[3]I_n[2]}{E_n -E_0} + \frac{2\mu^2 vt}{3mR^4} 
\sum_{n=2}^{\infty} \frac{(I_n[3])^2}{E_n -E_0} + O(\varepsilon^{3}).
 \end{eqnarray}
So, the magnetic monopole external field leads to the 
appearance of a nonzero electric dipole moment of the hydrogen atom 
 which, as expected, is proportional to the product of the charges 
of the monopole and the electron.

To calculate the radial integrals $I_n[k]$ we use the expression 
for the radial functions
\begin{equation}                        \label{radial}
R_{n1} = \frac{2}{3}\frac{{\sqrt{n(n^2-1)}}}{n^3} 
r e^{-r/n} {_1F_1}(-n+2,4;\frac{2r}{n});\quad R_{10} = 2 e^{-r}
\end{equation}
where  ${_1F_1}(-N,a,x)$ is the standard
confluent hypergeometric function. 
Taking  account its  well-known  properties  
\cite{14},  it  is  easy  to  calculate  those  integrals in explicit form:
\begin{equation}
I_n[2] = \int\limits_{0}^{\infty} dr r^2 R_{n1}R_{10} =
2^3 \frac {n{\sqrt{n(n^2-1)}}}{(n+1)^4} {_2F_1} (-n+2,4,4,\frac{2}{n+1});
\end{equation}
\begin{equation}
I_n[3] = \int\limits_{0}^{\infty} dr r^3 R_{n1}R_{10} =
2^5 \frac {n^2{\sqrt{n(n^2-1)}}}{(n+1)^5} {_2F_1} (-n+2,5,4,\frac{2}{n+1});
\end{equation}
Thus
\begin{eqnarray}
\sum_{n=2}^{\infty}
 \frac{I_n[3]I_n[2]}{E_n -E_0}&=&\frac{2^9}{me^2Q^2} 
\sum_{n=2}^{\infty} \frac{n^6}{(n+1)^9}~{_2F_1} 
(-n+2,5,4,\frac{2}{n+1}) {_2F_1} (-n+2,4,4,\frac{2}{n+1} );
\nonumber\\
\sum_{n=2}^{\infty} \frac{(I_n[3])^2}{E_n -E_0} &=&
\frac{2^{11}}{me^2Q^2} \sum_{n=2}^{\infty} 
\frac{n^7}{(n+1)^{10}}~ \biggl({_2F_1} (-n+2,5,4,\frac{2}{n+1})\biggr)^2.
\end{eqnarray}
After some numerical calculations one can obtain in the first order of
perturbation theory
\begin{equation}    \label{e-dip}
\left| <{\tilde \Psi}_0|x-iy|{\tilde \Psi}_0>\right| = 
 \left| <{\tilde \Psi}_0|x+iy|{\tilde \Psi}_0>\right| 
\approx \frac{b}{(meQ)^2} 
\frac{\mu }{R^2} + O(\varepsilon^{2}) 
= 2b \frac{\Delta E_0}{E_0} +O(\varepsilon^{2}).
\end{equation}
Note, that the appearance of the 
electric dipole moment (\ref{e-dip}) of the hydrogen atom 
in the monopole presence is not connected with the 
well known extra angular momentum in the charge-monopole system.   
 
\bigskip

{\bf Acknowledgements}
\medskip

I acknowledge numerous conversations with E.A.Tolkachev and  L.M.Tomilchik. 
I am also very indebted to Prof.\ Per Osland  for
fruitful discussions and hospitality at the University of Bergen
where this work has been
completed.
This research  has been supported by 
the Alexander von Humboldt Foundation in the framework 
of the Europa Fellowship.

\end{document}